\documentclass{elsart}
\usepackage{graphics}
\newcommand{\be}{\begin{equation}}
\newcommand{\ee}{\end{equation}}
\newcommand{\ZZ}{\mathbf{Z}}
\newcommand{\NN}{\mathbf{N}}
\begin{document}
\begin{frontmatter}
\title{A New Class of Automata Networks}
\author[UIC,Saclay]{Nino Boccara},
\author[UIC]{Henryk Fuk\'s} and
\author[UIC]{Servie Geurten}
\address[UIC]{Department of Physics, University of Illinois, Chicago, 
IL 60607-7059, USA}
\address[Saclay]{DRECAM-SPEC, CE Saclay, 91191 Gif-sur-Yvette Cedex, 
France}
\begin{abstract}
A new class of automata networks is defined. Their evolution rules are determined
by a probability measure $p$ on the set of all integers $\ZZ$ and an indicator
function $I_A$ on the interval $[0,1]$. It is shown that any cellular automaton
rule can be represented by a (nonunique) rule formulated in terms of a pair
$(p,I_A)$. This new class of automata networks contains discrete systems which
are not cellular automata. For a given $p$, a metric can be defined on the space
of all rules which induces a metric on the space of all cellular automata rules.
\end{abstract}
\end{frontmatter}

\section{The Evolution Operator}
Systems which consist of a large number of simple identical elements evolving in
time according to simple rules often exhibit a complex behavior as a result of
the cooperative effect of their components. Automata network are models of
such systems. They consist \cite{gm90} of a graph with a
discrete  variable at each vertex which evolves in discrete time steps
according to a definite rule involving the values of neighbouring vertex
variables. The vertex variables may be updated sequentially or synchronously.
More formally, automata networks may be defined as follows:

Let $G=(V,E)$ be a graph, where $V$ is a set of vertices and $E$ a set of edges.
Each edge joins two vertices not necessarily distinct.  An automata network,
defined on $V$, is a triple $\left(G,Q, \{f_i|i \in V \} \right)$, where $G$ is a graph on
$V$, $Q$ a finite set  of states and $f_i\colon Q^{|U_i|}\to Q$ a mapping,
called the transition rule associated to vertex $i$. $U_i=\{j\in V|\{j,i\}\in
E\}$  is the neighbourhood of $i$, i.e., the set of vertices connected to $i$, 
and $|U_i|$ denotes the number of vertices belonging to $U_i$. If $|U_i|$ is
finite, the rule is local. For the automata networks considered in this paper,
$|U_i|$ is infinite, however it will be shown that, if certain conditions are
fulfilled, even if the number of vertices in the neighborhood of a vertex
is infinite, the rule may be local. Cellular automata (CA) are  particular
automata networks in which the set of vertices $V$ is the set of all
integers $\ZZ^d$, where $d$ is called the space dimensionality, and the rule is
translationally invariant, local and applied synchronously. 

Since the class of automata networks to be described is a generalization of
CA, to fix the notations, we define CA as follows: 
Let $s:\ZZ\times\NN\mapsto\{0,1\}$ be a function that satisfies the equation
\begin{eqnarray}
(\forall i\in\ZZ)\ (\forall t\in\NN)\quad
 s(i,t+1)=f\big(s(i-r,t), \nonumber \\
s(i-r+1,t),\ldots,s(i+r,t)\big),
\end{eqnarray}
where $\NN$ is the set of nonnegative integers and $\ZZ$ the set of all
integers. Such a discrete dynamical system is a two-state one-dimensional
CA. The mapping $f : \{0,1\}^{2r+1}\to\{0,1\}$ is the rule, and the
positive integer $r$ is the radius of the rule. More general rules could be
site- and/or time-dependent and/or involve different left and right radii. The
function  $S_t:i\mapsto s(i,t)$ is the state of the CA at time $t$. ${\mathcal
S}=\{0,1\}^{\ZZ}$ is the state space. An element of the state space is also
called a configuration. Since the state $S_{t+1}$ at time $t+1$ is entirely
determined by the state $S_t$ at time $t$ and the rule $f$, there exists a
unique mapping  $F_f:{\mathcal S}\to{\mathcal S}$ such that $S_{t+1}=F_f(S_t)$. $F_f$,
which is the evolution operator, is also referred to as the CA rule.

CA have been widely used to model complex systems in which the local character of
the rule plays an essential role \cite{ftw84,w86,mbvb89,g90,bgmp93}. Nevertheless, not all complex
systems exhibit purely local interactions. Consider, for example, diffusion of
innovations in a social system. For a given individual, deciding to buy, say,
her first computer is based on the word of mouth from, more or less, closed
neighbors as her family members, friends and colleagues as well as mass-media
communications. It is clear that a CA rule cannot correctly model this type of
system. The new class of automata networks which described in this paper might
be useful to model such systems. 

The evolution operator $F_{p,A}$ of our new class of automata networks is
defined in terms of a probability measure $p$ on $\ZZ$ and an indicator
function $I_A$ on $[0,1]$. Then, we shall show that, for any CA rule $f$ , we
can find such an evolution operator, that is, a measure $p$ and an indicator
$I_A$, that emulates the CA rule. 

Let $S_t$ be the state of the system at time $t$, that is, $S_t:i\mapsto s(i,t)$,
and put 
\be
\sigma(i,t)=\sum_{n=-\infty}^\infty s(i+n,t)p(n), \label{sigdef}
\ee
where $p$ is a given probability measure on $\ZZ$, that is, a nonnegative
function on the set of all integers such that 
\be
\sum_{n=-\infty}^\infty p(n)=1. \label{norm}
\ee
For all $i\in\ZZ$ and $t\in\NN$, $\sigma(i,t)\in[0,1]$. The state $S_{t+1}$ of
the system at time $t+1$ is then determined by the function 
\be
i\mapsto s(i,t+1)=I_A\left(\sigma(i,t)\right)
=I_A\left(\sum_{n=-\infty}^\infty s(i+n,t)p(n)\right), 
\ee
where $I_A$ is a given indicator function on $[0,1]$, that is, a function such
that, for all $x\in[0,1]$,
\be
I_A(x)=\left\{ \begin{array}{ll} 
              1, & \mbox{if $x\in A\subset [0,1]$,} \\
              0, & \mbox{otherwise.}
               \end{array} 
\right. 
\ee
Since the state $S_{t+1}$ at time $t+1$ is entirely determined by the state
$S_t$ at time $t$ and the measure $p$ and the subset $A$ of $[0,1]$, there exists
a unique mapping  $F_{p,A}:{\mathcal S}\to{\mathcal S}$ such that 
\begin{eqnarray*}
S_{t+1}=F_{p,A}(S_t).
\end{eqnarray*}
$F_{p,A}$ will be referred to as the evolution operator.

For a given probability measure $p$ and a given indicator function $I_A$,
the operator $F_{p,A}$ is entirely determined. That is, the word ``probability
measure'' should not be misleading: the evolution operator $F_{p,A}$ is
deterministic. Probabilistic evolution operators could also be defined. If, for
instance, we replace $I_A$ by $XI_A$, where $X$ is a Bernoulli random variable
taking values on $\{0,1\}$, the resulting evolution operator would be
probabilistic.
\begin{thm}
Let $F_f$ be the evolution operator on $\mathcal S$
associated to a CA rule $f$; then, there exists an evolution operator $F_{p,A}$
such that, for any configuration $x\in{\mathcal S}$, $F_{p,A}(x)=F_f(x)$.
\end{thm}
To simplify the proof, we only consider translation-invariant symmetric CA rules,
that is, rules $f$ such that, for all $i\in\ZZ$ and all $(2r+1)$-block
$B(i,r)=\{x(i-r),x(i-r+1),\ldots,x(i+r)\}$ ($r>0$), 
\begin{eqnarray}
f\big(x(i-r),x(i-r+1),\ldots,x(i+r)\big)= \nonumber \\
f\big(x(i+r),x(i+r-1),\ldots,x(i-r)\big).
\end{eqnarray}
The generalization to nonsymmetric rules is straightforward. Since $f$ is
symmetric, we shall verify that we may assume that $p$ is even, that is, for
all $n\in\ZZ$, $p(-n)=p(n)$. The set of configurations with prescribed values
at a finite number of sites is called a cylinder set. To each $(2r+1)$-block
$B(i,r)$ corresponds a cylinder set denoted $C(i,r)$. Since the rule is
translation-invariant, it is sufficient to consider $i=0$.

For any configuration $x:n\mapsto x(n)$ belonging to the cylinder set $C(0,r)$,
the set of all numbers $\xi\big(C(0,r)\big)$ defined by 
\be
\xi\big(C(0,r)\big)=\sum_{n=-\infty}^\infty x(n)p(n)
\ee
belongs to the subinterval 
$[\xi_{\rm min}\big(C(0,r)\big),\xi_{\rm max}\big(C(0,r)\big)]$ of $[0,1]$
(called a $C$-interval in what follows) such that 
\be
\xi_{\rm min}\big(C(0,r)\big)=\sum_{n=-r}^r x(n)p(n)
=x(0)p(0)+\sum_{n=1}^r\big(x(-n)+x(n)\big)p(n)
\ee
and
\be
\xi_{\rm max}\big(C(0,r)\big)=\xi_{\rm min}\big(C(0,r)\big)
+2\sum_{n=r+1}^\infty p(n), 
\ee
where we have taken into account that $p(-n)=p(n)$.

Since, for $n\ne 0$, $\xi_{\rm min}\big(C(0,r)\big)$ depends on $x(-n)+x(n)$ and
not on $x(-n)$ and $x(n)$ separately, there are only $2\times 3^r$ different
$\xi_{\rm min}\big(C(0,r)\big)$ (and as many $C$-intervals). It is, therefore,
more convenient to label them $\xi_{\rm min}(\nu)$, where $\nu$ is an integer in
$\{0,1,\ldots,2\times 3^r-1\}$ defined by
\be
\nu=x(0)3^r+\sum_{k=1}^r\big(x(-k)+x(k)\big)3^{r-k}.
\ee
The corresponding $\xi_{\rm min}(\nu)$ is then given by
\be
\xi_{\rm min}(\nu)=x(0)p(0)+\sum_{k=1}^r\big(x(-k)+x(k)\big)p(k).
\ee

If we want to define $2^{2\times 3^r}$ different operators $F_{p,A}$
representing the $2^{2\times 3^r}$ different symmetric CA rules $f$ of radius
$r$, we have first to find a probability measure $p$ such that the $2\times 3^r$
$C$-intervals are disjoint. Then, according to the rule $f$ to be represented, we
will choose a subset $A$ of $[0,1]$ such that some of these $C$-intervals are
strictly included in $A$ whereas the others have an empty intersection with $A$.
The only problem is, therefore, to find the conditions to be satisfied by $p$.

The sequence $\{\xi_{\rm min}(\nu)\mid \nu=0,1,\ldots,2\times 3^r-1\}$ is totally
ordered for increasing values of $\nu$ if, and only if, the $r+1$ conditions
\begin{eqnarray}
p(r-k)>2p(r-k+1)+2p(r-k+2)\cdots+2p(r) \label{cond1} \\
 (k=0,1,\ldots,r)  \nonumber
\end{eqnarray}
are satisfied, with the convention: $p(m)=0$ if $m<0$. 

It is easy to verify that, for $\nu=0,1,\ldots,2\times 3^r-1$, the difference 
$\xi_{\rm min}(\nu+1)-\xi_{\rm min}(\nu)$ is always of the form
\begin{eqnarray}
p(r-k)-2p(r-k+1)-2p(r-k+2)\cdots-2p(r) \\
 (k=0,1,\ldots,r). \nonumber
\end{eqnarray}
Therefore, if all these differences are greater than 
\begin{eqnarray*}
\xi_{\rm max}\big(C(0,r)\big)-\xi_{\rm min}\big(C(0,r)\big)=
2\sum_{n=r+1}^\infty p(n),
\end{eqnarray*}
all the intervals 
\begin{eqnarray*}
\Big[\xi_{\rm min}(\nu),\; \xi_{\rm min}(\nu)+2\sum_{n=r+1}^\infty p(n)\Big]
\end{eqnarray*}
will be disjoint. For this to be the case, the $r+1$ conditions
\be
p(r-k)>2\sum_{n=r-k+1}^\infty p(n)\quad (k=0,1,\ldots,r). \label{cond2}
\ee
should be satisfied. Since $p$ satisfies (\ref{norm}), Conditions (\ref{cond2})
 are identical to Conditions (\ref{cond1}).
 
\section{Examples}
Here are two examples of probability measures $p$.

{\it Example 1}.
If, for all $n\in\ZZ$,
\be
p(n)=\tanh(\half \lambda)\exp(-\lambda |n|),
\ee
where $\lambda>0$, it is easy to verify that Conditions (\ref{cond1}) are satisfied for
all positive values of $r$ if $\lambda>\log 3$. That is, choosing a subset $A$
of $[0,1]$ we can represent, in this case, any symmetric CA rule. For a given CA
rule $p$ and $A$ are clearly not unique. 

If we take $\lambda=1.2$, the corresponding $C$-intervals are approximately
\begin{eqnarray*}
{[}\xi_{\rm min}(0),\xi_{\rm max}(0)] & = & [0,0.1395]      \\
{[}\xi_{\rm min}(1),\xi_{\rm max}(1)] & = & [0.1618,0.3012] \\
{[}\xi_{\rm min}(2),\xi_{\rm max}(2)] & = & [0.3235,0.4630] \\
{[}\xi_{\rm min}(3),\xi_{\rm max}(3)] & = & [0.5370,0.6765] \\
{[}\xi_{\rm min}(4),\xi_{\rm max}(4)] & = & [0.6988,0.8382] \\
{[}\xi_{\rm min}(5),\xi_{\rm max}(5)] & = & [0.8606,1].
\end{eqnarray*}
And to obtain Rule 18 \cite{w83} defined by
\begin{eqnarray*}
f(x_1,x_2,x_3)= \left\{ 
              \begin{array}{ll}
             1, & \mbox{if $(x_1,x_2,x_3)=(0,0,1)$ or $(1,0,0)$,} \\
             0, & \mbox{otherwise,}
               \end{array}
\right.
\end{eqnarray*}
we may choose $A=[0.15,0.31]$.

{\it Example 2}.
If, for all $n\in\ZZ$,
\be
p(n)=\frac{(1+|n|)^{-\alpha}}
{2\zeta(\alpha)-1},
\ee
where $\alpha>1$ and $\zeta$ is the zeta function, then, for a radius equal to
$r_0$, there exist a threshold value $\alpha_0$ such that 
Conditions (\ref{cond1}) are satisfied for all $r\le r_0$ if 
$\alpha\ge\alpha_0$ where $\alpha_0$ is the solution of the equation 
\begin{eqnarray*}
(1+r_0)^{-\alpha_0}-2\sum_{n=r_0+1}^\infty (1+n)^{-\alpha_0}=0.
\end{eqnarray*}
The table below give some approximate threshold values.
\begin{eqnarray*}
\begin{array}{cccccccc}
r_0 & 0 & 1 & 2 & 3 & 10 & 20 & 100 \\
\alpha_0\; & 2.18529\; & 3.28994\; &4.39061\; &5.49023\; &13.1824\; &24.169\; &112.058
\end{array} 
\end{eqnarray*}
$\alpha_o$ is approximately a linear function of $r_0$.

For $\alpha=3.5$, the corresponding $C$-intervals are approximately
\begin{eqnarray*}
{[}\xi_{\rm min}(0),\xi_{\rm max}(0)]&=&[0,0612]\\
{[}\xi_{\rm min}(1),\xi_{\rm max}(1)]&=&[0.0705,0.1317]\\
{[}\xi_{\rm min}(2),\xi_{\rm max}(2)]&=&[0.1410,0.2022]\\
{[}\xi_{\rm min}(3),\xi_{\rm max}(3)]&=&[0.7978,0.8590]\\
{[}\xi_{\rm min}(4),\xi_{\rm max}(4)]&=&[0.8683,0.9295]\\
{[}\xi_{\rm min}(5),\xi_{\rm max}(5)]&=&[0.9388,1].
\end{eqnarray*}
And to obtain Rule 18, we may choose, in this case, $A=[0.062,0.133]$.

If some $C$-intervals are not disjoint but are either strictly included in $A$
or have an empty intersection with $A$, certain CA rules could be represented,
but not all CA rules of a given radius. To represent nonsymmetric CA rules we
should consider noneven functions $p$.

An operator $F_{p,A}$ will not, in general, represent a CA rule in the following
two cases.
\begin{enumerate} 
\item[(i)] If $p$ is such that for a given $r$ the $C$-intervals are
disjoint but $A$ has a nonempty intersection with some $C$ intervals without,
however, strictly including these intervals.
\item[(ii)] If $p$ is such that for a given $r$ the $C$-intervals are not
disjoint.
\end{enumerate}

In the first case, the spatio-temporal pattern will look like the pattern
of a CA whose evolution is governed by a mixed rule. But, a mixed rule being
local, the velocity of propagation of a local perturbation cannot be greater than
$r$, whereas in this case the value of the $\sigma$-variable defined by
 (\ref{sigdef})
depends on the whole configuration and not only on a finite block. Note that
the nonlocal character of the evolution operator $F_{p,A}$ comes essentially
from the definition of the indicator function since if we consider a measure
$p$ whose moments are all defined (as in Example 1) the range of the
interaction between the different sites is clearly finite whatever the precise
meaning we give to this range.  As a consequence of the nonlocal character of
the rule in this case the velocity of propagation of a local perturbation may be
time-dependent and grow to infinity.

In the second case, the spatio-temporal pattern may look like a CA pattern, but
no finite radius can be defined. As an example, consider the
exponential probability measure, and choose $\lambda=1$. The $C$-intervals are
not disjoint and the resulting evolution operator is not a CA rule. For
two different initial configurations, the corresponding spatio-temporal
patterns of such an operator are represented in Figures 1a and 1b.
\begin{figure}
\begin{center}
a) \scalebox{0.7}{\includegraphics{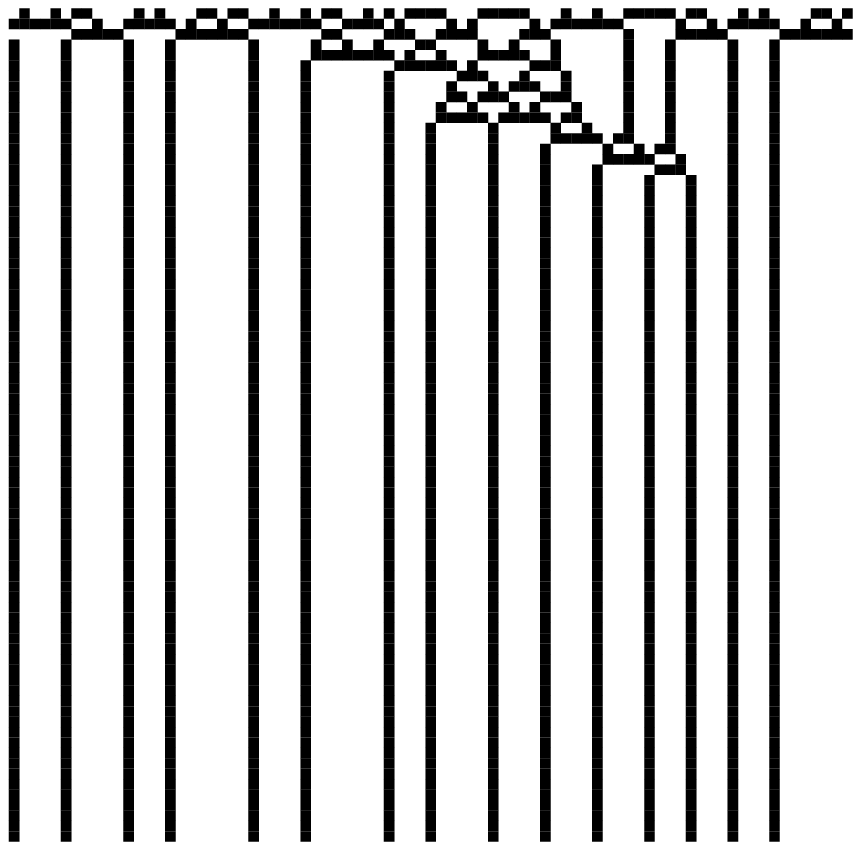} }
b) \scalebox{0.7}{\includegraphics{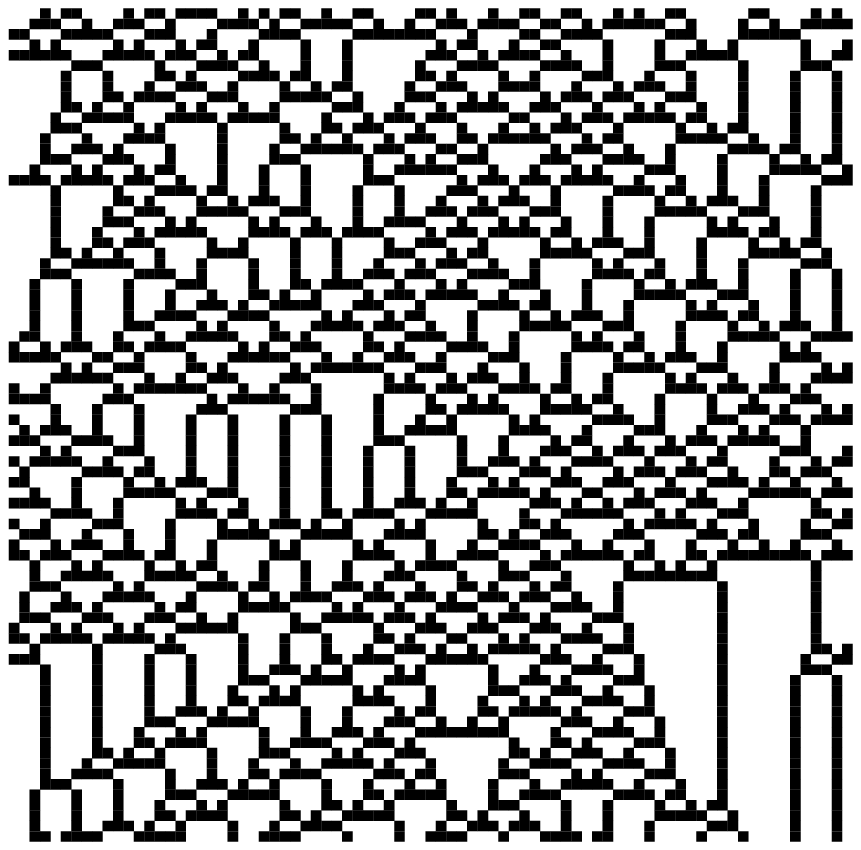} }
\end{center}
\caption{Spatiotemporal pattern for $A=[0.197,0.600]$ and $\lambda=1$ for two
different initial configurations.}
\end{figure}
While Pattern
1a is reminiscent of a class-2 CA, Pattern 1b looks more like class-3. Wolfram
\cite{w83} classification is, therefore, not relevant for rules of this
type (failure of Wolfram classification has already been noticed even for
standard CA, as reported in \cite{hc92} and \cite{dlmw95}).

CA have been widely used to model systems, say in epidemiology 
\cite{bc92a,bc92b,bco93,brr94} or in ecology \cite{brr94}, in
which the local character of the interaction is an essential feature. But, there
are cases in which the interactions are both local and nonlocal as for the
diffusion of innovations \cite{mmb90,r95} which can be
influenced by two types of communication channels: mass media and
interpersonal. The new class of discrete dynamical systems we have defined may
be well adapted to model such a situation. If, for instance, we consider a
probability measure $p$ such that $p(n)$ is proportional to
\be
{\e}^{-\lambda_1 |n| }+a \left({\e}^{-\lambda_2|n-n_0|}+
{\e}^{-\lambda_2|n+n_0|}\right),
\ee
 where $\lambda_1$, $\lambda_2$ and $a$ are positive constants, and $n_0$ is a
positive integer. $\lambda_1$ and $\lambda_2$ are inversely proportional to the
range of the nearby and distant sources of information, $a$ measures the
relative weight of the distant source compared to the nearby one and $n_0$ is
the location of the distant source.  A model of diffusion of innovations
employing this idea is currently investigated and a detailed report will be
published elsewhere.

\section{Metric}
For a given probability measure $p$, let $F$ be the probability distribution of
the random variable $\Sigma:{\mathcal S}\to [0,1]$ defined by
\be
\Sigma=\sum_{n=-\infty}^\infty X(n)p(n),
\ee
where $\{X(n)\mid n\in\ZZ\}$ is a doubly infinite sequence of equally
distributed Bernoulli random variables such that
\be
(\forall n\in\ZZ)\qquad P(X(n)=0)=P(X(n)=1)=\half.
\ee
\begin{thm}
If $\lambda>\log(3)$, the random variable $\Sigma$ is
singular.
\end{thm}
That is, for $\lambda>\log 3$, the distribution function $F$ of $\Sigma$ is
continuous but has no density. The derivative of $F$ is almost everywhere equal
to zero, and $F$ increases on a Cantor-like set of zero Lebesgue measure (see
Figures 2a and 2b). 
\begin{figure}
a) \include{fig2a} 
b) \include{fig2b} 
\caption{Distribution function for a) $\lambda=1.3$ and 
 b) $\lambda=2$ .}
\end{figure}
This implies that, if an evolution operator is selected at
random by selecting a subset $A$ of $[0,1]$ at random, then, with probability
one, the corresponding rule will be a CA rule.

As for the usual triadic Cantor set, the set on which $\Sigma$ is defined can
be constructed by removing step by step ``forbidden intervals.'' 
\begin{eqnarray*}
\mbox{1st step}\;\, & & \\
\sigma &\notin& \Big]2\sum_{n=1}^\infty p(n),p(0)\Big[\\
\mbox{2nd step} & & \\
\sigma &\notin& \Big]2\sum_{n=2}^\infty p(n),p(1)\Big[\\
& & \bigcup\Big]p(1)+2\sum_{n=2}^\infty p(n),2p(1)\Big[\\
& & \bigcup\Big]p(0)+2\sum_{n=2}^\infty p(n),p(0)+p(1)\Big[\\
& & \bigcup\Big]p(0)+p(1)+2\sum_{n=2}^\infty p(n),p(0)+2p(1)\Big[
\end{eqnarray*}
etc. For a given radius $r$, the sum $m(r)$ of the Lebesgue measures of the removed
intervals is
\begin{eqnarray}
m(r)&=&\sum_{\nu=1}^{2\times 3^r-1}(\xi_{\rm min}(\nu)-\xi_{\rm max}
(\nu-1))     \\
&=&(p(0)+2p(1)+\cdots+2p(r))\nonumber \\
& & +(2\times 3^r-1)\,2\sum_{k=r+1}^\infty p(k). \nonumber
\end{eqnarray}
Since
\be
\lim_{r\to\infty}(2\times 3^r-1)\,2\sum_{k=r+1}^\infty p(k)=0.
\ee
it follows that
\be
\lim_{r\to\infty} m(r)=1.
\ee

This result could be made more intuitive in the following way. If we put
$\lambda=\log b$,  
\be
\Sigma=\frac{b-1}{b+1}\left(X(0)+\sum_{n=1}^\infty\frac{X(-n)+X(n)}
{b^n}\right). \label{Sigdef}
\ee
Since, for all positive integers $n$, the sum of the Bernoulli
random variables $X(-n)$ and $X(n)$ is either equal to 0, 1 or 2, the
 summation over $n$ in (\ref{Sigdef}) represents a random variable whose  values 
are expressed in base $b$ ($b$ is not necessarily an integer). If $b>3$, the set
of all possible values of $\Sigma$ is, therefore, a Cantor-like set.

For $b=3$, the random variable $\Sigma$ is given by
\be
\Sigma=\frac{1}{2}\left(X(0)+\sum_{n=1}^\infty\frac{X(-n)+X(n)}{3^n}\right).
\ee
$X(-n)+X(n)$ being a binomial random variable such that
\begin{eqnarray*}
P\big(X(-n)+X(n)=0\big)&=& \quart \\
P\big(X(-n)+X(n)=1\big)&=& \half  \\
P\big(X(-n)+X(n)=2\big)&=& \quart
\end{eqnarray*}
$\Sigma$ is absolutely continuous but not uniformly distributed on $[0,1]$ (see
Figure 3),
\begin{figure}
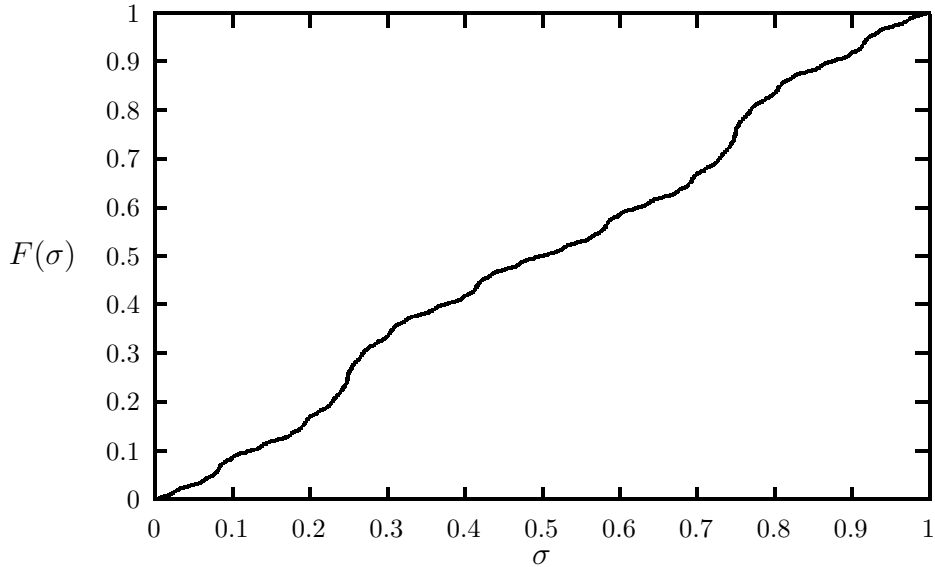

\include{fig3}  
\caption{Distribution function for $\lambda=\log{3}$.}
\end{figure}
and the set of all possible values of $\Sigma$ has a nontrivial
Hausdorff dimension $d_H$, which, in this very particular case, can be exactly
determined (see, for example, \cite{b95}). It is found that
\be
d_H=\frac{3}{2}\frac{\log 2}{\log 3}=0.946395\ldots.
\ee

For a fixed probability measure $p$ we define the distance between two
evolution
operators ${\bf F}_{p,A_1}$ and ${\bf F}_{p,A_2}$ by
\be
d({\bf F}_{p,A_1},{\bf F}_{p,A_1})
=\int_0^1|I_{A_1}(\sigma)-I_{A_2}(\sigma)|\,dF(\sigma)
=\int_{A_1\triangle A_2}dF(\sigma),
\ee
where $A_1\triangle A_2$ denotes the symmetrical difference between $A_1$ and
$A_2$, that is, the set of all numbers $\sigma\in[0,1]$ that belong
to $A_1$ XOR
$A_2$.

Note that $d({\bf F}_{p,A_1},{\bf F}_{p,A_1})=0$ does not imply $A_1=A_2$.
This means that two evolution operators may be different but the corresponding
dynamics will be identical. Therefore, we could, instead of considering
evolution operators we could consider classes of evolution operators. Two
evolution operators would belong to the same class if their distance is equal to
zero. This notion of distance may be useful to study sequences of, say, cellular
automaton rules and study if such sequences converge in the topology defined by
this metric. This topic will be developed elsewhere.
\begin{thm}
If $\lambda>\log(3)$, the distance between two CA rules
does not depend upon $\lambda$.
\end{thm}
This result follows from the fact that the subset $A$, characterizing a CA rule,
is a reunion of disjoint intervals whose end points belong to removed intervals.
It can even be proved that the distance will not depend upon the form of $p$
provided that $p$ is such that Conditions (\ref{cond1}) are satisfied for
all values of $r$ less or equal than the larger radius of the rules under
consideration. For example, the distance between Rule 18 (defined above) and Rule
22 defined by 
\begin{eqnarray*}
f(x_1,x_2,x_3)= \left\{ \begin{array}{ll}
1, &  \mbox{if $(x_1,x_2,x_3)=(0,0,1)$ or $(1,0,0)$ or $(0,1,0)$,} \\ 
0, & \mbox{otherwise,}
\end{array}
\right.
\end{eqnarray*}
is
\be
d({\bf F}_{18},{\bf F}_{22})= \int_{\{0,1,0\}}dF(\sigma)=\frac{1}{8}.
\ee
It is equal to the probability of selecting $\{0,1,0\}$ among the 8 equally
probable blocks of radius one.

\section{Conclusion}
We have described a new class of automata networks whose
evolution operators are defined in terms of a probability measure $p$ and an
indicator function $I_A$, where $A$ is a subset of the interval $[0,1]$. This
class contains all cellular automaton rules, and other rules which have no finite
radius. For a given $p$, we have defined a metric on the space of these
new rules which induces a metric on the space of all CA rules that does not
depend upon the particular $p$. In the case of evolution operators which are
not CA rules, the Wolfram classification into 4 classes does not seem to be
relevant since we may obtain spatio-temporal patterns reminiscent of two
different classes by changing the initial configuration. The parameters
characterizing the evolution operators of these new automata networks can be
modified continuously, and, as a consequence, starting from a given initial
configuration, we can, say, pass continuously from a class-3 to a class-2 CA.
Finally these new rules could be useful to model systems in which the
interactions between the different elements are short- and long-range, like in
the diffusion of innovations.

\end{document}